\documentclass[aps,11pt,showpacs,showkeys,a4paper]{revtex4}
\usepackage{graphicx}
\usepackage{dcolumn}
\usepackage{bm}
\begin{document}

\title{Application of the Tool--Narayanaswamy--Moynihan model to the study of the $\alpha$ relaxation by thermally stimulated depolarization currents}

\author{J. Sellar\`es}
\email{jordi.sellares@upc.edu}
\author{J.C. Ca\~nadas}
\author{J.A. Diego}
\author{M. Mudarra}
\author{J. Belana}
\affiliation{Departament de F\'{\i}sica i Enginyeria Nuclear,
Universitat Polit\`ecnica de Catalunya (Campus de Terrassa),
c.~Colom 1, E-08222 Terrassa, Catalonia, Spain}

\date{\today}

\begin{abstract}

The dielectric $\alpha$ relaxation is the dielectric manifestation of the glass transition. In spite of this fact, the more commonly used models, the Arrhenius model and the Williams--Landel--Ferry model, do not take into account the structural state of the system to modelize this relaxation. In thermally stimulated discharge current (TSDC) experiments, the sample is out of equilibrium during most of the discharge ramp. Not surprisingly, the capability of these models to represent the data points at temperatures well below the glass transition temperature is very poor. To overcome this limitation, we have used the Tool--Narayanaswamy-Moynihan (TNM) model, that takes into account the structural state of the system, to modelize TSDC data. Although it is mostly applied to calorimetric and volumetric experiments we show how it can be employed on dielectric data. The numerical results support the applicability of the model and suggest how the dielectric and the structural relaxation times may be related. The TNM model turns out to be physically sound since the modelization of dielectric data gives also a reasonable structural kinetics of the system. The qualitative differences between the TNM model and the equilibrium models for different TSDC experiments is discussed. Experimental data coming from TSDC experiments where the thermal history of the sample is changed do not provide additional evidence but nevertheless are compatible with the parameter values that were obtained in the fits to the TNM model.

\end{abstract}

\pacs{81.40.Tv, 77.22.Gm, 77.22.Ej}

\keywords{structural relaxation; dielectric relaxation; thermally stimulated depolarization current}

\maketitle

\section{Introduction}
\label{intro}

A glass is a material that behaves mechanically like a solid but has a
disordered structure. In fact, glasses can be considered as overcooled
liquids that have acquired amorphous rigidity. The most common way of
making a glass is by cooling the material from the melt fast enough to
avoid crystallization (although many polymers do not crystallize at any
cooling rate). If a liquid is cooled in such a way, molecules will
rearrange progressively more slowly. Eventually the structure of the
material will not be able to reach the equilibrium conformation fast
enough to follow the cooling rate. This falling out of equilibrium is
called the glass transition \cite{kenna}. From this point of view, the
glass transition would not be a true phase transition because all physical
magnitudes would change in a continuous way. It has been suggested that it
exists an underlying phase transition \cite{kenna} but anyway the glass
transition occurs across a narrow range of temperatures and therefore we
can define a glass transition temperature ($T_g$). Due to the kinetic
nature of the glass transition, $T_g$ depends on the cooling rate, but
this dependence turns out to be rather weak \cite{canvitg}. The study of
the glassy state properties has received a great amount of
attention \cite{kenna} and it is nowadays a very active subject of
research, both about systems close to $T_g$ \cite{revista1} or clearly
below this temperature \cite{revista2}.

In a wide sense, an electret is a sample of dielectric material that produces a nearly permanent external electric field. This field results from permanent ordering of molecular dipoles or displacement and trapping of free--charge. The polarization and the relaxation processes of an electret are intimately related to the properties of the material. Therefore, an study of these processes can provide useful information about the material. A technique in wide use to perform such studies is thermally stimulated depolarization currents (TSDC) \cite{sessler}.

A classical TSDC study can be described in the following way: the sample is heated up to a polarization temperature ($T_p$) and polarized between two electrodes for a polarization time ($t_p$) by a polarizing electric field ($E_p$). Then it is cooled down to a temperature, called storage temperature ($T_s$), low enough so that depolarization takes place at a very low rate. If conventional polarization is employed, the polarizing electric field is applied not only during $t_p$ but also during a large portion of the cooling ramp. Alternatively, in window polarization (WP) \cite{descobridor} the electric field is switched off at a point of the cooling ramp placed a few degrees below $T_p$ or even at the beginning of the cooling ramp.

Once the sample is at temperature $T_s$, it is kept at this
temperature for a storage time ($t_s$) and then it is heated at a
constant rate while it is shortcircuited through an electrometer. The obtained
displacement current is recorded as a function of temperature. In the thermogram
that results, the relaxation processes can be seen as current peaks that
form the TSDC spectrum. Electret behavior and TSDC have been widely
described in the literature \cite{sessler,bolena}.

Despite their experimental simplicity, TSDC spectra of conventionally polarized electrets are not easy to interpret. On one hand, different mechanisms can contribute to the thermogram, such as space charge relaxation and dielectric relaxation. On the other hand, these mechanisms may not be elementary (i.e. they can not be described in terms of a single relaxation time).

It is possible to resolve a complex TSDC spectrum into elementary spectra using the WP technique \cite{descobridor,sauer,teyssedre}. For this reason, WP is a useful technique to study relaxation mechanisms. As a consequence of employing WP only part of the mechanism remains activated. If the thermal window is narrow enough, the part of the mechanism that is activated will behave approximately in an elementary way. These spectra are well described by a single relaxation time.

One of the relaxations that can be detected by TSDC is the $\alpha$ relaxation,
which is the dielectric manifestation of the glass transition. In TSDC data, the
$\alpha$ peak has its maximum at the dynamic value of $T_g$
\cite{quimics1}. This value is the one at which really takes place the
glass transition during the heating ramp and usually lies some degrees
above the static value of $T_g$. Unlike in other relaxations
\cite{other1,other2,other3}, the Arrhenius law
\begin{equation}
\tau(T)=\tau_0 \exp \left(\frac{E_a}{RT}\right)
\label{arrheniuslaw}
\end{equation}
does not work properly when the system is close to the glass transition
because cooperative phenomena are involved \cite{honey}. The temperature
dependence is better described by a Vogel--Tammann--Fulcher (VTF) type
expression
\begin{equation}
\tau(T)= \tau_0 \exp \left[ \frac{E_w}{R(T - T_\infty)} \right] ,
\label{vogelfulcher}
\end{equation}
where $T_\infty$ is the temperature at which molecules would ``freeze''
and the relaxation time of the system would become infinite. It can be shown
that this model is equivalent to the Williams--Landel--Ferry equation. In
either case, $\tau(T)$ has no longer its initial physical meaning, a
characteristic time that represents the return of the excited mechanism to
equilibrium. This is also due to the cooperative character of the
mechanism. For this reason the pre--exponential factor has values that are not
easy to interpretate in mechanical terms. In spite of this fact, it is
still a valid empirical parameter to characterize the relaxation
\cite{sauer,colette}.

Both expressions can be used to modelize TSDC data. Nevertheless, an
important point is often missed: in eqs.~\ref{arrheniuslaw} and
\ref{vogelfulcher}, $\tau(T)$ represents the relaxation time in
equilibrium at a temperature $T$. To record a TSDC scan of the $\alpha$
relaxation, the sample has to be cooled well below $T_g$ and then heated
over this temperature in a short amount of time. As a result, the
configuration of the molecules that form the sample is far from
equilibrium during most of the experiment. The departure from equilibrium
depends on the kinetics of the system and the design of the experiment.

There are a number of ways to extend these equations to describe the
non-equilibrium glassy state \cite{hodge,maromo1,maromo2}. Many of them
are phenomenological theories that have been often tested with
calorimetric and dilatometric data \cite{hodge,maromo1}. Other models
follow a more first--principles approach \cite{bisquert,coupling}.
Occasionally, models that represent glassy behavior have been used to
modelize TSDC data. For example, the coupling model \cite{coupling} has
been used for this purpose \cite{vander}.

Polymethyl methacrylate (PMMA) is a polar, glassy and transparent polymer
with multiple industrial applications. Its most common application is as
a replacement for window glass because of its optimal chemical, mechanical
and optical properties. Since this material has been intensively studied
and the relaxations it can undergo have a rather simple behavior, it is
also appropriate as a test field for new techniques such as space charge
distribution measurements \cite{sessler} or second harmonic generation in
guest-host systems \cite{sessler}.

The aim of this work is to demonstrate how the Tool--Narayanaswamy--Moynihan (TNM) model can be used to modelize data from the $\alpha$ relaxation of PMMA obtained by TSDC. Since this model takes into account the structure of the system, this approach highlights the relationship between the dielectric $\alpha$ relaxation, the structural relaxation and the glass transition. The physical implications and the coherence of the model will be discussed in order to evaluate its applicability to dielectric phenomena.

\section{Experimental}
\label{experimental}


The experimental setup for TSDC measurements is composed of a custom made convection oven placed inside a cold bath at $-20$~$^\circ$C. The oven is controlled by an Eurotherm $2416$ temperature programmer. Inside, there is a measurement cell with two electrodes that are applied to the sample in plane--parallel configuration. The current was measured by a Keithley $6512$ electrometer. Temperature and current data were collected by an A/D converter card and processed afterwards.


PMMA samples were cut from a commercial sheet of 1.5~mm thickness. The thickness of the samples was reduced to $0.75$~mm by mechanical procedures. Circular aluminum electrodes of $1$~cm of diameter were deposited in vacuum on both sides of the sample. Smaller samples were also cut for differential scanning calorimetry (DSC) experiments.


DSC experiments were performed to determine the glass transition temperature at several heating rates. We used a Mettler TC 11 thermoanalyser equipped with a Mettler-20 differential scanning calorimeter module. DSC curves were obtained from 15~mg samples sealed in aluminum pans. The heating rate of the scans was 20~$^\circ$C/min, 10~$^\circ$C/min, 5~$^\circ$C/min and 2.5~$^\circ$C/min. From a these results it can be extrapolated that $T_g \approx 110$~$^\circ$C/min. The dynamic value at $2.5$~$^\circ$C/min is $T_g \approx 115$~$^\circ$C/min.


The TSDC experiments carried out follow the scheme represented in fig.~\ref{tsdc}. The values taken by the parameters shown in fig.~\ref{tsdc} are listed in table~\ref{experiments}. All the ramps have a rate of $2.5$~$^\circ$C/min. Experiments start well above $T_g$ in order to clear the influence from any previous thermal history and to begin the experiment with the sample at equilibrium state. The sample is annealed for an annealing time ($t_a$) at an annealing temperature ($T_a$).

Next, the sample is polarized. In all the presented experiments $T_p=T_a$. Most of them use $T_p=95$~$^\circ$C. This polarization temperature was chosen in order to isolate as much as possible the $\alpha$ peak from the space charge ($\rho$) peak, but without being too far from the optimal polarization temperature ($T_{po}$) of the $\alpha$ relaxation. In previous experiments with the same sample the $T_{po}$ was found to be $107.5$~$^\circ$C. A polarization potential of $800$~V and a polarization time of $900$~s ($15$~min) were employed in all the experiments.

Once the polarization time is over, the field is switched off and a cooling ramp begins. This ramp continues until $T_s$ is attained. After $t_s$ the sample is short-circuited through an electrometer and heated again while the thermally stimulated discharge current is recorded.

\section{Results and discussion}
\label{resdis}

The TSDC thermogram that we will use to illustrate the application of the TNM model is shown in fig.~\ref{general} and has been obtained with the experiment \#1 described in table~\ref{experiments}. The peak corresponds to the dielectric $\alpha$ relaxation. To analyze these data we will assume the Debye model. According to this model, the relaxation process presents an ideal--viscous rotational friction characterized by a single relaxation time ($\tau$)
\begin{equation}
\frac{dP}{dt} = \frac{1}{\tau} (P_{eq} - P).
\label{viscous_friction}
\end{equation}
This magnitude is strongly dependent on the temperature, so we will refer to it as $\tau(T)$.

The Bucci method \cite{bucci_method} is the framework in which the evaluation of the experimental value of $\tau(T)$ has been obtained. This method is based in the fact that the Debye model given by eq.~\ref{viscous_friction} is based in first--order kinetics. From this fact and other general considerations the expression
\begin{equation}
\tau(T) = \frac{1}{v I(T)}\int_{T}^{T_\infty} I(T') dT'
\label{temps_de_relaxacio_experimental}
\end{equation}
can be deduced. In this equation $v$ is the constant heating rate and $I(T)$ is the intensity of the recorded TSDC current. Given $\tau(T)$ it is possible to recover the intensity (up to an arbitrary factor) from the relaxation time, using the formula
\begin{equation}
I(T) \propto \exp \left\{ - \frac{1}{v} \int_{T_i}^{T} \frac{dT'}{\tau(T')}
- \ln \left[ \tau(T) \right] \right\} .
\label{procesdeprimerordrealreves}
\end{equation}

The characteristics of this magnitude are best appreciated in a $\ln(\tau)$ vs. $1/T$ plot. To visualize better the correspondence between the features in the intensity plot and in the relaxation time plot, the intensity has also been plotted in fig.~\ref{general} using a reciprocal temperature scale.

These data has been fitted to several models by $\chi^2$ minimization. The routines used are described by other authors \cite{recipes}. A point that deserves some attention is for which magnitude the maximum likeness should be achieved. A perfect model should give the same results independently of the magnitude used for the minimization, but this is obviously not our case. Although there are many possibilities, we have decided to perform our fits for two of them: $I$ (as in \cite{vander,vander2}) and $\ln(\tau)$ (as in \cite{grenet}). The parameters resulting from the $I$ fit, aside from giving better results for this magnitude, also represent better the behavior of the points placed near the maximum. Instead, parameters coming from $\ln(\tau)$ fits represent better the overall behavior of the experimental data. Anyway, the difference between the plots that have been computed with both sets of parameters gives a qualitative measurement of how appropriate is the model. This idea will be used throughout to assess the models.

In many works (see for example \cite{treball_pei}) the data fitted corresponds to the temperature range where $I(T)$ is greater than $I(T_{max})/2$. In fig.~\ref{general} that would mean fitting the data points that lay between $T'_1 = 95.5$~$^\circ$C and $T_2 = 110.5$~$^\circ$C. In this region $\ln[\tau(1/T)]$ has almost a linear behavior. Moreover, it has a slight concave curvature, so it is also possible to apply the WLF model.

In fig.~\ref{arrhenius}a we compare the fits to an Arrhenius model with $I$ data. The intensity plot that has been fitted to $I$ data (point--dash--dashed line) represents fairly well the data placed between $T'_1$ and $T_2$. On the other hand, the plot fitted to $\ln(\tau)$ data (point--dashed line) is very close to the previous one. The same happens in fig.~\ref{arrhenius}b that represents $\ln(\tau)$ data and plots obtained with the same parameters as in fig.~\ref{arrhenius}a. The plots obtained fitting to $\ln(\tau)$ data and to $I$ data are very similar. The parameter values obtained from these fits can be seen in table~\ref{arrheniusfits}

In fig.~\ref{wlf} we illustrate how the WLF model is even more successful than the Arrhenius model in this range of temperatures. Both kinds of fits, to $I$ data and to $\ln(\tau)$ data, give almost the same parameter values (see table~\ref{wlffits}). The representation of both solutions is almost the same, either in fig.~\ref{wlf}a or in fig.~\ref{wlf}b.

We can see, then, that the equilibrium models, Arrhenius and WLF, represent finely the behavior of the system in a narrow temperature range around the dynamic $T_g$. Definitively, they are appropriate for this purpose, with the WLF even better than the Arrhenius model in this case. Let's now go on to consider what happens when a wider range of temperatures is considered. Instead of taking $T'_1$ as the lower limit, we will fit the data from $T_1 = 70$~$^\circ$C. In this way we include the data points that form the foot of the $\alpha$ peak but we still avoid the contribution that could arise from the $\beta$ relaxation. It is not possible to enlarge the temperature range on the high temperature end, that we will keep at $T_2 = 110.5$~$^\circ$C. The reasons are that the accuracy of the integral in eq.~\ref{temps_de_relaxacio_experimental} becomes too low and also that the contribution from space charge relaxation turns out to be too noticeable in the results.

We will point out, in first place, that the WLF model can no longer be applied. There is an inflexion point near $T'_1$ in such a way that most of the $\ln[\tau(1/T)]$ curve has now convex curvature and the WLF model would give unphysical results, with $T_\infty > T_g$. Fig.~\ref{arrhenius} shows the results for the Arrhenius model (see numerical values in table~\ref{arrheniusfits}). It is clear that fits to $I$ data and to $\ln(\tau)$ data give plots that differ much more than in the previous case. This is a clear sign that in this range of temperatures the model is no longer appropriate.

Results obtained fitting $\ln(\tau)$ data do not represent well neither the relaxation time nor the intensity, as seen in fig.~\ref{arrhenius}. The plots obtained fitting $I$ data placed between $T_1$ and $T_2$ are very similar to the ones obtained fitting data points between $T'_1$ and $T_2$ because, as mentioned earlier, this procedure takes more into account the points that are placed closer to the peak maximum. Nevertheless, in fig.~\ref{arrhenius}a it can be seen that the values of the intensity that lay between $T_1$ and $T'_1$ cannot be modelized properly.

A classical explanation would be that, even when WP is employed, several parts of the mechanism are polarized, giving rise to a distribution of relaxation times. According to this explanation, the thermogram can not really be treated as if it were elementary.

In fact, it is true that the relaxation is distributed to some extent. Some experimental results can not be explained otherwise. Nevertheless, there are reasons to think that, even in this case, the Arrhenius model does not provide an appropriate basis to express the TSDC intensity. A broad distribution of relaxation times would be needed to modelize the region between $T_1$ and $T'_1$. Instead, the data points between $T'_1$ and $T_2$ can be modelized quite well using a single relaxation time. This suggests a narrow distribution, which is more likely in WP experiments. There would be no contradiction if the intensity at the beginning of the $\alpha$ peak corresponding to an elementary relaxation would rise up not as suddenly as in the Arrhenius model. Otherwise, usual distribution functions such as Fuoss--Kirkwood \cite{fuo41} or Havriliak--Negami \cite{hav67} will not be able to give good results. Therefore, the problem remains more or less the same as before.

There is a chance that this failure is due to the out--of--equilibrium state of the material during the TSDC ramp. In the following lines we will discuss how the TNM model can be applied to provide a more realistic profile of the TSDC intensity. To lighten up the discussion we will keep the formalism at the elementary relaxation level although it should be emphasized that the TNM model is not incompatible with the existence of a distribution of relaxation times.

The TNM \cite{orignm} phenomenological model has been used mostly to modelize calorimetric and dilatometric data. Within this model, the relaxation time depends not only on the present temperature but also on the thermal history of the sample. We will denote this relaxation time as $\tau_s$ and we will not suppose yet that it is identical or otherwise related to the dielectric relaxation time $\tau$.

In the TNM model the structure dependence is taken into account through
the fictive temperature ($T_f$). The fictive temperature of a
non--equilibrium system is equal to the temperature of an equilibrium
system that has the same structure. The separation between temperature and
fictive temperature dependence is introduced through a non--linearity
parameter $x$ ($0 \leq x \leq 1$). The structural relaxation time is given
by
\begin{equation}
\ln(\tau_s) = \ln(\tau_{s0}) + \frac{x E_a^*}{RT} + \frac{(1-x)
E_a^*}{RT_f},
\label{NM}
\end{equation}
where $E_a^*$ is the apparent activation energy. The TNM model is reduced
to the Arrhenius model given by eq.~\ref{arrheniuslaw} either when the system
is at equilibrium or in the limit when the non--linearity parameter is
equal to $1$.

The kinetics of $T_f$ is determined by the first--order equation
\begin{equation}
\frac{dT_f}{dt} = - \frac{T_f - T}{\tau_s(T,T_f)}.
\label{dinNM}
\end{equation}
In other words, we are assuming the same type of kinetics for both the
dielectric and the structural relaxation.

Eqs.~\ref{NM} and \ref{dinNM} are coupled. To solve the system, we define
a reduced time
\begin{equation}
z(t)= \int_0^t \exp\left[-\frac{x E_a^*}{RT(t')}-\frac{(1-x)
E_a^*}{RT_f(t')}\right] dt'.
\label{tempsreduit}
\end{equation}
In terms of the reduced time, the solution of eqs.~\ref{NM} and
\ref{dinNM} reads
\begin{eqnarray}
T_f(t) - T(t) &=& \phi\left[z(t)\right] \left[ T_f(0) - T(0) \right] \nonumber
 \\
& & -\int_0^t \frac{dT(t')}{dt'} \phi\left[z(t)-z(t')\right] dt'.
\label{solucionm}
\end{eqnarray}
We have written the solution also in terms of the response function of the
system
\begin{equation}
\phi(t)\equiv\exp\left(-\frac{t}{\tau_{s0}}\right).
\label{respostanm}
\end{equation}
This function determines the temporal behavior of the system in response
to an external perturbation that has been applied such as an electrical
field, a stress or a deformation.

If $T(0)>T_g$ the first term of the RHS of eq.~\ref{solucionm} vanishes
because the fictive temperature and the temperature are equal above $T_g$.
If the system is below $T_g$ at the beginning of the experiment, there is
no easy way to know the initial fictive temperature. $T(t)$ is a known function
that reproduces the temperature from the beginning of the experiment to the end.
It is possible to calculate $\tau_s(T)$ at any time of the experiment
although only the $\tau_s(T)$ that corresponds to the TSDC ramp
can be compared to the experimental one.

Eqs.~\ref{tempsreduit} and \ref{solucionm} are as coupled as eqs.~\ref{NM}
and \ref{dinNM} and the integrals therein have to be evaluated
numerically. To accomplish this, the time variable $t$ is discretized and
the integrals are calculated in the following iterative way. First, the
fictive temperature of the previous time step is used to evaluate
eq.~\ref{tempsreduit}. The resulting reduced time is introduced in
eq.~\ref{solucionm} to calculate a closer approximation to the fictive
temperature at the present time step. If the difference with the previous
approximation to the fictive temperature is not small enough, the process
is iterated.

The time step taken has been of $5$~s and Simpson's rule has been chosen
to evaluate the integrals. Under these conditions, the convergence of the
procedure that has been described is very fast and is achieved within few
iterations, even if the precision required is of four significant digits.

Up to this point, we know how to calculate the relaxation time of the structural relaxation ($\tau_s$) at any time for a
given set of parameters ($E_a^*$, $\tau_{s0}$, and $x$) and a thermal history $T(t)$. Now we should assume a relationship between $\tau_s$ and the relaxation time of the dielectric relaxation ($\tau$).

The simplest one is that both relaxation times are the same
\begin{equation}
\tau(T)=\tau_s(T),
\label{hipotesi1}
\end{equation}
but in our case it does not work. This can be seen in fig.~\ref{nm3p} (numerical results in table~\ref{nmfits}). The model fits a little bit better to the experimental points because it has one more parameter and contains the Arrhenius model as a particular case but the same qualitative problems that we found for the Arrhenius model remain. In fig~\ref{nm3p}a, the parameters obtained fitting the TNM model intensity data give an almost identical curve to the Arrhenius model. Only the relaxation times close to the maximum are well modelized with these parameters (fig~\ref{nm3p}a), just as happened using the same procedure with the Arrhenius model. The results obtained fitting the model to relaxation time data are as bad as in the Arrhenius model. The parameters obtained in this way are not able to reproduce the relaxation time data (fig~\ref{nm3p}b), and obviously they perform even worse when they are used to reproduce the intensity (fig~\ref{nm3p}b). To sum up, this hypothesis has more or less the same drawbacks as the Arrhenius model in spite of being much more involved.

In terms of simplicity, the next relationship to consider would be a proportionality between the relaxation times
\begin{equation}
\tau(T) \propto \tau_s(T).
\label{hipotesi2}
\end{equation}
This hypothesis implies the introduction of an additional parameter. This parameter can be expressed as another pre--exponential factor
\begin{equation}
\tau_{d0} \equiv \tau_{s0} \frac{\tau(T)}{\tau_s(T)}.
\label{factor_pre_exponencial_dielectric}
\end{equation}
As within the previous hypothesis, the Arrhenius model is attained when $x = 1$, without the need of additional conditions.

The results are shown in fig.~\ref{nm4p} (see table~\ref{nmfits} for numerical values). The ability to match the experimental data is much better, either when we fit intensity data (continuous line of fig.~\ref{nm4p}a) or relaxation time data (dashed line of fig.~\ref{nm4p}b). In the case of the intensity, the model is able to modelize the points placed between $T_1$ and $T'_1$, unlike previous models that could not cope with those points. The model can reproduce finely the position of the inflexion point and the curvature of each zone in relaxation time data. Another clear sign that the model performs better than the previous ones is that the curves obtained fitting $I$ and $\ln(\tau)$ are very close, either in fig.~\ref{nm4p}a or in fig.~\ref{nm4p}b.

Although this model is more accurate than the previous ones, some limitations can be noticed. The match of the theoretical curve with experimental data becomes worse as we consider points that are placed further below $T_g$. This phenomena is reasonable since the TNM model is obtained by extending an equilibrium model (the Arrhenius model) and therefore it should be less valid as the system attains structural states further from equilibrium. The difference between the parameter values obtained fitting to intensity data and to relaxation time data is due to this disagreement.

From the previous discussion it should be clear that the TNM model, together with the proportionality hypothesis, is able to fit TSDC experimental data with more accuracy than the Arrhenius and the WLF models. By itself this is not enough to consider that the TNM model provides an appropriate representation of the dielectric $\alpha$ relaxation. Better fits could be expected anyway due to the presence of more fitting parameters. On the following lines we will discuss the kinetics of the structure of the system within the TNM model to gain insight on its physical meaning and it will be checked that the kinetic behavior predicted by the model is the one that should be expected according to our current knowledge on physical aging.

In fig.~\ref{dynamics} we represent the evolution of the difference between $T_f$ and $T$, for two different initial deviations. This magnitude can be considered as the deviation from structural equilibrium of the system. The time origin has been taken at the beginning of the cooling ramp that takes place after the first isotherm.

The effect of the cooling ramp is to increase dramatically the deviation from equilibrium. The dotted lines have been introduced as a guide for the eye and represent slopes of $2.5$~$^\circ$C/min in the scale of the $T_f-T$ plot. This is the rate of all the ramps in fig.~\ref{tsdc}. From the plot follows that during the cooling ramp $T_f$ tends to attain a constant value (about $90$~$^\circ$C for experiment \#2, according to our calculations) and therefore the deviation will tend to grow at a rate equal to the one of the cooling ramp.

The deviation does not decrease significantly during the storage time. For this reason during the TSDC ramp the deviation has its maximum value at the beginning of the ramp. Since the TNM is just a non--equilibrium extension of the Arrhenius model, this is the part of the ramp that will be harder to modelize. The dotted line allows us to see that, until $t_1$, $T_f$ remains more or less the same. Before this time, the change in $T_f-T$ is due mainly to the variation of $T$. After $t_1$, $T_f$ begins to change very quickly. At $t_2$ the system has almost reached the structural equilibrium state. According to this behavior, the temperatures at those two times should correspond to the static and the dynamic $T_g$. $T(t_1)=103$~$^\circ$C should, therefore, be roughly the static $T_g$. DSC results mentioned earlier suggest that this estimation is a few degrees beneath the actual value. $T(t_2)=115$~$^\circ$C is approximately equal to the dynamic $T_g$ obtained by DSC. It should be taken into account that $T_g$ is not an input of the model. The only way that the model has to obtain the structural kinetics of the system is from the $I$ data. For this reason parameters resulting from a fit to an experiment with $T_p=T_{po}$ have been employed, since these parameters ought to be more representative of the behavior of the $\alpha$ relaxation as a whole. Nevertheless, it is remarkable that the picture of the glass transition is so realistic as the one that has been obtained.
%

Now that we have seen that the results from the TNM model give a reasonable structural kinetics of the system, we are able to discuss which features characteristic of the TNM model or incompatible with it can be found in data coming from usual TSDC experiments. An important feature of the model is that it takes into account the structural state of the sample.
%
%
For this reason, the results given by the model should depend on the thermal history. Nevertheless, the existence of a distribution of relaxation times can also yield this effect. In fact, a complete model should include some kind of relaxation time distribution modelization to be able to predict results coming from a wide range of experiments. Nevertheless, there are several ways to implement such modelization and a complete discussion lies out of the scope of this work. Instead, we will discuss several experiments and explain why their results are compatible with a TNM relaxation time even when the TNM model by itself may not be able to explain all the features observed in the experimental data, without making further assumptions.

A usual way to obtain thermal history dependent results is changing the sub $T_g$ annealing time. The TNM model predicts that the structural state of the system at the end of a cooling ramp will depend almost exclusively on the cooling rate. As a consequence, the structural state of the sample during the TSDC ramp will vary in the same way in all the experiments. No change should be expected on the $\ln(\tau)$ plot due to this cause. This does not mean that such changes are not possible. In fact, polarization takes place when the sample is at different structural states. In the same way that the total polarization of the sample decreases with annealing time, the part of the mechanism that is polarized can also change. In fig.~\ref{aging} we present experimental results using annealing times between $3$~min and $243$~min. From these experimental data we can establish that in the case of PMMA the $\ln(\tau)$ plot does not change significantly during the first hour of annealing at $95$~$^\circ$. We can infer that we are polarizing the same part of the mechanism using this range of annealing times. The $\ln(\tau)$ plots change slightly for larger annealing times due to a small shift towards greater temperatures of the $I$ plot. This does not mean that the TNM model is no longer valid but rather that we are not polarizing the same part of the mechanism.



From the previous lines follows that there are only three parameters that we can change in order to observe a characteristic TNM behavior in the experimental data: storage temperature, cooling rate and heating rate of the TSDC ramp.

The easiest way to obtain different values of $T_f-T$ at the beginning of the TSDC ramp is employing several storage temperatures. Nevertheless, the structural state of the system will not vary very much neither at the end of the cooling ramp, during the storage time nor at the beginning of the TSDC ramp. This happens because temperatures suitables for storage are so far from $T_g$ that the structure of the system is almost frozen. The predicted intensity plots according to the TNM model can be seen at fig.~\ref{diposit}. The hypothetical parameters that have been employed have been chosen in order to make the variation of the plots more noticeable. The foot of the $\alpha$ peak tends to grow for higher storage temperatures while the peak maximum exhibits a slight shift towards lower temperatures. This behavior is at a first glance different from the one expected from the Arrhenius or the WLF models, that predict no change in the data. Anyway, the TNM model can give a behavior similar to the one of the equilibrium models. The key parameter is the structural pre--exponential factor. For values of $\tau_{s0}$ less than $10^{-19}$~s, the difference between the plots becomes very difficult to appreciate and smaller than the experimental resolution. Experimental $I$ plots obtained with two storage temperatures are presented in fig.~\ref{diposit2}. The experimental data presents no differences between the two thermograms but this result is compatible with the parameters resulting from the fit to $I$ data. Fits to $I$ data may be more representative because the performance of the model is worse for structural states further from equilibrium and fits to $I$ data tend to overestimate the influence of the data points placed near $T_{max}$, that are not as far from structural equilibrium as the other ones.

Calculations assuming $\tau_{s0} \approx 10^{-19}$~s show that few changes should be expected using different cooling rates. In fact, experimental results (not reported in this paper) show a displacement of the $\alpha$ peak towards higher temperatures as slower cooling rates are employed. This effect is explained easily if the existence of a distribution of relaxation times is assumed. It is not incompatible with a TNM behavior but it will certainly mask any characteristic TNM effect that could be observed.

Finally, the heating rate of the TSDC ramp has a strong and well--known influence on the $I$ data plots. The shift to higher temperatures of the $\alpha$ peak when the heating rate is increased does not depend on the model and can be explained using eq.~\ref{procesdeprimerordrealreves} exclusively. The $\ln(\tau)$ plot will show almost no change, according to all the models.

%
%

\section{Conclusions}

In this work we have shown how the TNM model can be used to modelize a more realistic shape of the TSDC intensity curve. Although the Arrhenius and the WLF models are the currently employed models in the literature, there are reasons to think that their description of the dielectric $\alpha$ relaxation is incomplete. The main one is that the structural state of the system should be taken into account in the modelization of a relaxation that is the dielectric manifestation of the glass transition.

An important question is which is the relationship between the structural relaxation time given by the TNM model and the dielectric relaxation time obtained by the analysis of TSDC data. Our calculations indicate that a proportionality relationship between these two quantities is a useful hypothesis. The numerical results obtained in this way are able modelize the $\alpha$ relaxation over a wider temperature range than the Arrhenius or the WLF models.

The application of the TNM model to dielectric data has its rough edges. The TNM model is, in fact, an extension of the Arrhenius model and therefore it is not surprising that its performance is better when the structural state of the system is not too far from equilibrium. As a consequence, the data points obtained at lower temperatures are not so well modelized as the ones placed nearer to the $T_g$ and the result of the fits depend somewhat on which is the magnitude for which the maximum likeness is searched.

Nevertheless there are also some promising features. The TNM model provides a good description of the relaxation over a wider range of temperatures than the equilibrium models. The modelization of the dielectric data provides also a realistic kinetics of the structure of the system, without the need of stating explicitly the glass transition temperature. The dependence of the available experimental data on the thermal history can be given by the TNM model if the pre--exponential factor of the structural relaxation time has values below $\approx 10^{-19}$. The parameter values obtained fitting the TNM model with an additional parameter to intensity data is compatible with the behavior observed in available experimental results.

All in all, the numerical results support the convenience of taking into account the out of equilibrium state of the system, although a really fine tuned experiment would be needed to find qualitative evidence of its influence.

This research is being continued along several lines. Other phenomenological models can be compared to the TNM model and may provide better results. The combination of phenomenological models with a distribution of relaxation times should give an improved description of the system. The use of other techniques such as dielectric electrical analysis can also give useful information. Finally, these new methods can be applied to review the classical studies on TSDC and physical aging.


\begin{acknowledgments}

This work has been partially supported by project {\em MAT 2001--2338--C02--01} ({\em MCT}, Spain).

\end{acknowledgments}


\newpage

\pagestyle{empty}

\begin{table}[h]

\caption{Parameters of the experiments that have been performed.}
\label{experiments}

\vspace{0.25cm}

\begin{center}
\begin{tabular}{cccccccc}\hline
Experiment \# & $T_a$~($^\circ$C) & $t_a$~(s) & $T_p$~($^\circ$C) & $t_p$~(s) & $V_p$~(V) & $T_s$~($^\circ$C) & $t_s$~(s) \\ \hline
1 & 95    & 1080  & 95    & 900 & 800 & 40 & 840 \\
2 & 107.5 & 180   & 107.5 & 900 & 800 & 30 & 840 \\
3 & 95    & 180   & 95    & 900 & 800 & 40 & 840 \\
4 & 95    & 1080  & 95    & 900 & 800 & 40 & 840 \\
5 & 95    & 1980  & 95    & 900 & 800 & 40 & 840 \\
6 & 95    & 3780  & 95    & 900 & 800 & 40 & 840 \\
7 & 95    & 7380  & 95    & 900 & 800 & 40 & 840 \\
8 & 95    & 14580 & 95    & 900 & 800 & 40 & 840 \\
9 & 95    & 1080  & 95    & 900 & 800 & 40 & 840 \\
10 & 95    & 1080 & 95    & 900 & 800 & 60 & 840 \\
\end{tabular}
\end{center}

\end{table}

\newpage

\begin{table}[h]

\caption{Results obtained from the fits to the Arrhenius model}
\label{arrheniusfits}

\vspace{0.25cm}

\begin{center}
\begin{tabular}{cccccc}\hline
Fit & Experiment \# & Variable & Range~($^\circ$C) & $E_a$~(eV) & $\tau_0$~(s)\\ \hline
(a) & 1 & $I$         & 95.5--110.5 & $1.96$ & $1.03 \times 10^{-24}$ \\
(b) & 1 & $\ln(\tau)$ & 95.5--110.5 & $1.69$ & $4.84 \times 10^{-21}$ \\
(c) & 1 & $I$         & 70.0--110.5 & $1.90$ & $6.21 \times 10^{-24}$ \\
(d) & 1 & $\ln(\tau)$ & 70.0--110.5 & $1.37$ & $1.13 \times 10^{-16}$ \\
\end{tabular}
\end{center}

\end{table}

\newpage

\begin{table}[h]

\caption{Results obtained from the fits to the WLF model}
\label{wlffits}

\vspace{0.25cm}

\begin{center}
\begin{tabular}{ccccccc}\hline
Fit & Experiment \# & Variable & Range~($^\circ$C) & $E_w$~(eV) & $\tau_0$~(s) & $T_{\infty}$~(K) \\ \hline
(e) & 1 & $I$         & 95.5--110.5 & $2.60 \times 10^{-2}$ & $2.70 \times 10^{-1}$ & $330$ \\
(f) & 1 & $\ln(\tau)$ & 95.5--110.5 & $2.60 \times 10^{-2}$ & $3.45 \times 10^{-1}$ & $329$ \\
\end{tabular}
\end{center}

\end{table}

\newpage

\begin{table}[h]

\caption{Results obtained from the fits to the TNM model}
\label{nmfits}

\vspace{0.25cm}

\begin{center}
\begin{tabular}{cccccccc}\hline
Fit & Experiment \# & Variable & Range~($^\circ$C) & $E_a$~(eV) & $x$ & $\tau_{s0}$~(s) & $\tau_{d0}$~(s) \\ \hline
(g) & 1 & $I$         & 70.0--110.5 & $2.08$ & $0.855$ & $2.05 \times 10^{-26}$ & $\tau_{s0}$ \\
(h) & 1 & $\ln(\tau)$ & 70.0--110.5 & $2.03$ & $0.607$ & $9.53 \times 10^{-26}$ & $\tau_{s0}$ \\
(i) & 1 & $I$         & 70.0--110.5 & $1.52$ & $0.609$ & $2.18 \times 10^{-20}$ & $1.36 \times 10^{-19}$ \\
(j) & 1 & $\ln(\tau)$ & 70.0--110.5 & $1.13$ & $0.391$ & $3.48 \times 10^{-15}$ & $2.09 \times 10^{-14}$ \\
(k) & 2 & $I$         & 97.5--117.5 & $1.91$ & $0.462$ & $2.27 \times 10^{-24}$ & $2.36 \times 10^{-23}$ \\
\end{tabular}
\end{center}

\end{table}

\newpage

\begin{table}[h]

\caption{Parameters of the experiments that have been simulated.}
\label{simulacions}

\vspace{0.25cm}

\begin{center}
\begin{tabular}{cccc}\hline
$E_a$~(eV) & $x$ & $\tau_{s0}$~(s) & $\tau_{d0}$~(s) \\ \hline
$1.12$ & $0.388$ & $2.48 \times 10^{-14}$ & $1.49 \times 10^{-13}$ \\
\end{tabular}
\begin{tabular}{cccccccc}\hline
Simulation \# & $T_a$~($^\circ$C) & $t_a$~(s) & $T_p$~($^\circ$C) & $t_p$~(s) & $V_p$~(V) & $T_s$~($^\circ$C) & $t_s$~(s) \\ \hline
1 & 95 & 1080 & 95 & 900 & 800 & 30 & 840 \\
2 & 95 & 1080 & 95 & 900 & 800 & 40 & 840 \\
3 & 95 & 1080 & 95 & 900 & 800 & 50 & 840 \\
4 & 95 & 1080 & 95 & 900 & 800 & 60 & 840 \\
\end{tabular}
\end{center}

\end{table}

\newpage

\setcounter{topnumber}{0}
\setcounter{bottomnumber}{0}
\setcounter{totalnumber}{15}
\renewcommand{\textfraction}{0}
\setlength{\intextsep}{0.1cm}

\clearpage

\newtheorem{peudefigura}{Figure}

\begin{peudefigura}
{\rm Outline of TSDC experiments. During the polarization time $800$~V are applied to the sample. All the experiments begin and end at 125~$^\circ$C or above.}
\label{tsdc}
\end{peudefigura}

\begin{peudefigura}
{\rm TSDC scan of the sample and its Bucci--Fieschi--Guidi plot. The following temperatures are marked in the figure: $T_1 = 70$~$^\circ$C, $T'_1 = 95.5$~$^\circ$C, $T_{max} = 104$~$^\circ$C and $T_2 = 110.5$~$^\circ$C.}
\label{general}
\end{peudefigura}

\begin{peudefigura}
{\rm Arrhenius model fits. The model values obtained are listed in table~\ref{arrheniusfits}. Fits performed between $T'_1$ and $T_2$: point--dash--dashed plots are fitted to $I$ data (fit a). Point--dashed plots are fitted to $\ln(\tau)$ data (fit b). Fits performed between $T_1$ and $T_2$: continuous line plots are fitted to $I$ data (fit c). Dashed line plots are fitted to $\ln(\tau)$ data (fit d).}
\label{arrhenius}
\end{peudefigura}

\begin{peudefigura}
{\rm WLF model fits. The model values obtained are listed in table~\ref{wlffits}. Continuous line plots are fitted to $I$ data (fit e). Dashed line plots are fitted to $\ln(\tau)$ data (fit f). Fits performed between $T'_1$ and $T_2$.}
\label{wlf}
\end{peudefigura}

\begin{peudefigura}
{\rm TNM model fits. The model values obtained are listed in table~\ref{nmfits}. Continuous line plots are fitted to $I$ data (fit g). Dashed line plots are fitted to $\ln(\tau)$ data (fit h). Fits performed between $T_1$ and $T_2$.}
\label{nm3p}
\end{peudefigura}

\begin{peudefigura}
{\rm TNM model fits with additional parameter. The model values obtained are listed in table~\ref{nmfits}. Continuous line plots are fitted to $I$ data (fit i). Dashed line plots are fitted to $\ln(\tau)$ (fit j). Fits performed between $T_1$ and $T_2$.}
\label{nm4p}
\end{peudefigura}

\begin{peudefigura}
{\rm Continuous line: calculated deviation from structural equilibrium ($T_f - T$) during experiments with different structural states at the beginning of the cooling ramp ($t=0$). From this point, the thermal history of is the same as in experiment \#2. Dashed line: Temperature of the sample ($T$). The following points are marked in the figure: $T(t_1)=103$~$^\circ$C and $T(t_2)=115$~$^\circ$C. The parameters employed are the same as obtained in fit k.}
\label{dynamics}
\end{peudefigura}

\begin{peudefigura}
{\rm $\ln(\tau)$ results obtained with different annealing times. The experiments are listed in table.~\ref{experiments}: \#3 ($t_a=180$~s, small circle), \#4 ($t_a=1080$~s, large circle), \#5 ($t_a=1980$~s, plus) and \#6 ($t_a=3780$~s, cross). The dashed line is the mean of the four data plots and is included as a guide for the eye. The $\ln(\tau)$ data were calculated from the experimental $I$ data presented in the small frame, that also includes data from experiments \#7 and \#8. Lower peaks correspond to larger annealing times.}
\label{aging}
\end{peudefigura}

\begin{peudefigura}
{\rm Dependence on storage temperature for the TNM model with additional parameter. The simulations are listed in table~\ref{simulacions}: \#1 ($T_s = 30$~$^\circ$C, continuous),  \#2 ($T_s = 40$~$^\circ$C, dashed), \#3 ($T_s = 50$~$^\circ$C, dot--dashed--dashed) and \#4 ($T_s = 60$~$^\circ$C, dot--dashed).}
\label{diposit}
\end{peudefigura}

\begin{peudefigura}
{\rm Experimental TSDC thermograms where two different storage temperatures have been employed. The experiments are listed in table~\ref{experiments}: \#9 ($T_s = 40$~$^\circ$C, continuous line) and \#10 ($T_s = 60$~$^\circ$C, circle)}
\label{diposit2}
\end{peudefigura}


\newcommand{\dibuix}[2]{%

\newpage

\pagestyle{empty}

\hbox{}\vspace{2cm}

\begin{center}
\includegraphics[width=10cm]{#1}
\end{center}

\vspace{3cm}

\noindent
{\bf Figure #2}

\noindent
J. Sellar\`es {\em et al.}, ``Application of the Tool--Narayanaswamy--Moynihan model to the study of dielectric $\alpha$ relaxation by thermally stimulated discharge currents''.

}

\dibuix{tsdc}{\ref{tsdc}}

\dibuix{general}{\ref{general}}

\dibuix{arrhenius-a}{\ref{arrhenius}a}

\dibuix{arrhenius-b}{\ref{arrhenius}b}

\dibuix{wlf-a}{\ref{wlf}a}

\dibuix{wlf-b}{\ref{wlf}b}

\dibuix{nm3p-a}{\ref{nm3p}a}

\dibuix{nm3p-b}{\ref{nm3p}b}

\dibuix{nm4p-a}{\ref{nm4p}a}

\dibuix{nm4p-b}{\ref{nm4p}b}

\dibuix{dynamics}{\ref{dynamics}}

\dibuix{aging}{\ref{aging}}

\dibuix{diposit}{\ref{diposit}}

\dibuix{diposit2}{\ref{diposit2}}





\end{document}